\documentclass[aps,prb,reprint,twocolumn,floatfix,superscriptaddress,nofootinbib]{revtex4}
\usepackage{graphicx}
\usepackage{amssymb}
\usepackage{amsmath}
\usepackage{lmodern}
\usepackage{color}
\usepackage{hyperref}
\usepackage{empheq}
\usepackage{epstopdf}
\begin{document}

\title
{Low-energy collective electronic excitations in a gold monolayer}


\author{Vyacheslav \,M.~Silkin}

\affiliation{Donostia International Physics Center, P. de Manuel Lardizabal 4, 20018 San
Sebasti\'an/Donostia, Spain}
\affiliation{Departamento de Pol\'{\i}meros y Materiales Avanzados: F\'{\i}sica, Qu\'{\i}mica y Tecnolog\'{\i}a, Facultad de Ciencias Qu\'{\i}micas, Universidad del Pa\'{\i}s Vasco UPV/EHU,
Apartado 1072, 20080 San Sebasti\'an/Donostia, Spain}
\affiliation{IKERBASQUE, Basque Foundation for Science, 48009 Bilbao, Spain}

\author{M.\,Yu.~Kagan}

\affiliation{National Research University Higher School of Economics, Myasnitskaya Ulitsa 20, 101000, Moscow, Russia}
\affiliation{P. L. Kapitza Institute for Physical Problems, Russian Academy of Sciences, Ulitsa Kosygina 2, 119334, Moscow, Russia}



\begin{abstract}

The low-energy plasmonic properties of a free-standing Au(111) monolayer, frequently referred to as goldene, are investigated by time-dependent density functional theory in the random phase approximation. In the evaluation of dielectric and loss functions the first-principle band structure characterized by a single $s-p$ energy band at the Fermi level is taken into account.
In this system with apparently simple band structure around the Fermi surface the calculations show the presence, besides the standard two-dimensional (2D) plasmon mode, at even lower energies of three anomalous branches in the momentum-energy region usually assigned to incoherent electron-hole pairs.
This is an unexpected result since the electronic system of the gold monolayer is characterized by a single energy band crossing the Fermi level.
The dispersion of these anomalous plasmonic branches is strongly anisotropic in-plane and gapless. A general feature is that at small in-plane momentum transfers the modes have a sound-like dispersion in certain momentum directions.
We rely the appearance of these anomalous modes to a notable deviation of the Fermi surface of a gold monolayer from a circular shape. As a result of its almost perfect hexagon shape, there is possibility to realize in a {\it single} electronic energy band a scenario of a multi-component electron system with the carrier groups moving with different Fermi velocities in certain momentum directions.
The found modes correspond to the out-of-phase charge oscillations of the carriers at different sides of the hexagonal Fermi surface.
Hence our work presents evidence how in a metal system with apparently simple band structure around the Fermi level its shape can result in conversion of single-particle charge excitations into collective charge excitations.

\end{abstract}

\maketitle

\section{Introduction}\label{sec1}

Since the discovery of graphene about two decades ago,\cite{noges04} two-dimensional materials have attracted great interest in many research fields.
One of the topics extensively studied by solid state community has been the plasmons in two-dimensional (2D) systems. Which kinds of charge excitations, namely incoherent electron-hole (e-h) pairs and a collective mode (plasmon) can be generated in the low-energy domain in 2D was known since the work on the electronic excitations in a 2D free-electron gas.\cite{stprl67} In contrast to three-dimensional bulk systems a 2D plasmon has a gapless excitation spectrum with a characteristic square-root dispersion at small momentum transfers. This picture describes well the plasmons in dilute electronic systems.\cite{anformp82} Moreover, it can still be appropriate for metallic systems. A first experimental observation of a 2D plasmon in metallic systems was realized using a Ag monolayer adsorbed on silicon surface.\cite{nahiprl01}
In doped graphene, in addition to two high-energy plasmons, may appear in the low-energy range, a 2D plasmon.\cite{wustnjp06,hwdaprb07} Its dispersion is essentially similar to that of a conventional 2D plasmon.\cite{stprl67,hwdaprb07,dahwprl09} Last years the similar collective electronic excitations have been studied in other 2D systems based on the layered materials.\cite{akboan26} This is facilitated by its relatively easily synthesis.\cite{nojipnas05,buhoacsn13,rebo2dm26}

On the other hand, the 2D electronic systems can be realized in the atomically thin metallic materials. One of the examples is the growth of the metallic adlayers on different substrates. However, usually the strong interaction with the substrate significantly modifies the intrinsic properties of the adlayers.\cite{liwaam15,anlifcse16,chfacr18}  At the same time, there exist attempts to synthesize even free standing atomically thin metal slabs.\cite{kash2dm25,bayaai26} For instance, those based on the gold atoms. One of the promising systems might be a gold monolayer.\cite{wawanl19,kashns24,patonc24,zhzhprb24}

Other practical rout consists in intercalation of adsorbed graphene with suitable atomic species. Usually this method is applied to decouple the graphene layer from the substrate. At the same time, sometimes this approach results in creation of 2D monolayers of a single intercalated element with intriguing properties.\cite{haviape18,yaviprl19,folinc20,rostprb20,maroprr22,bapumt22,luvejvst26}
For instance, intercalation of the gold atoms between graphene and the underlying silicon carbide crystal allows to synthesize a metallic system electronically very close to a free-standing Au monolayer.\cite{folinc20} Upon initial intercalation of the gold atoms, a single Au atomic layer presents a semiconducting electronic structure. However,  when the number of the intercalated gold atomic layers increases to two, a transition to a metallic state occurs.\cite{folinc20} In a recent publication it was shown that measured metallic Au band dispersion is very close to the calculated band structure of a free-standing gold monolayer.\cite{byroarx26}

Recently the experimental observation of the 2D plasmon in the intercalated metallic Au(111) monolayer was achieved.\cite{byroarx26} The plasmon dispersion measured at small momentum transfers by scanning near-field optical microscopy (SNOM) is in a good agreement with the theoretical model predictions based on the band structure calculations.

Motivated by this study, here we realize first principle calculations of the dielectric function and the excitation spectra of a free-standing Au(111) monolayer in the framework of time-dependent density functional theory in the random phase approximation. Our main interest is to reveal how the warping of the $s-p$ energy band around the Fermi level influences the low-energy dielectric screening in this system. Our calculations show that the low-energy screening properties of the gold monolayer are in qualitative difference to the free-electron gas (FEG) theory.\cite{stprl67} Whereas the conventional 2D plasmon is not influenced notably by the shape of the Fermi surface, the intra-band e-h continuum experiences dramatic modifications. In particular, we find that significant spectral weight is transferred from the e-h's to novel collective modes having peculiar gap-less dispersion and presenting strong in-plane anisotropic behavior. We trace how these modes disperse with momentum and inter-relations among them.

\section{Methods and Calculation Details}\label{Methods and computational details}

A free-standing Au(111) monolayer was considered in a super-cell geometry where the gold atomic sheets are separated by vacuum intervals corresponding to 8 bulk interlayer distances. In the plane, the gold atoms are arranged in a hexagonal structure like the (111) face of the fcc lattice. The chosen in-plane lattice constant of 2.874 \AA \, is very close to the bulk experimental value of 2.860 \AA.\cite{wobam10}

Within the linear response approach framework, in the two-dimensional (2D) reciprocal space the matrix of non-interacting density response function $\chi^o_{{\bf G}{\bf G}'}({\bf q},\omega)$ is expressed as
\begin{eqnarray}
\chi^o_{{\bf G}{\bf G}'}({\bf q},\omega) &=& \frac{2}{S}\sum^{\rm 2BZ}_{\bf k} \sum^{\rm occ}_{n} \sum^{\rm unocc}_{n'}
\frac{f_{n{\bf k}}-f_{n'{\bf k}+{\bf q}}}{\varepsilon_{n{\bf k}}-\varepsilon_{n'{\bf k}+{\bf q}}+(\omega+{\rm i}\eta)} \nonumber \\
&\times&\langle{\psi_{n{\bf k}}} |e^{-{\rm i}({\bf q}+{\bf G}){\bf r}}|
\psi_{n'{\bf k}+{\bf q}}\rangle \nonumber \\
&\times&\langle{\psi_{n'{\bf k}+{\bf q}}} |e^{{\rm i}({\bf q}+{\bf G}'){\bf r}}|\psi_{n{\bf k}}\rangle.
\label{chi0}
\end{eqnarray}
Here ${\bf G}$ and ${\bf G}'$ are 2D reciprocal lattice vectors and the vector ${\bf q}$ is limited to the 2BZ. The factor 2 accounts for the spin and $S$ is the 2D unit cell area. At zero temperature the Fermi occupation factor $f_{n{\bf k}}$ is represented by a step function, and $\eta$ is an infinitesimal.
$\varepsilon_{n{\bf k}}$ and $\psi_{n{\bf k}}$ are the Kohn-Sham energy eigenvalues and wave functions for the $n$-th band at 2D wave vector ${\bf k}$, respectively. We set the number of the $\bf G$ vectors to 51. In the summation we included all the valence occupied (occ) and unoccupied (unocc) states with energies up to 20 eV above the Fermi level. In such a way, both the intra- and inter-band transitions are taken into account on the same footing. A Monkhorst-Pack 288x288 grid of ${\bf k}$ vectors was used in the 2BZ sampling.\cite{mopaprb76}

To obtain the self-consistent electronic structure, the Kohn-Sham equations were solved employing our own band structure computer code.\cite{sichf84}
For description of the electron-ion interaction a Troullier-Martins norm-conserving pseudopotential was used.\cite{trmaprb91}
A value of the kinetic cutoff of the plane-wave basis was chosen to be 40 Ry. The self-consistent procedure was performed over a 24x24 ${\bf k}$ mesh in the 2D Brillouin zone (2BZ).

In practical calculations, one should choose some finite value for $\eta$ in Eq. (\ref{chi0}). However this procedure introduces some artificial broadening in all the features appearing in the calculated excitation spectra. Sometime it is useful, but in the present case we are interested in obtaining the intrinsic properties of the damping processes in a system. Therefore, to obtain $\chi^o$ without introduction of this artificial broadening, we use an equivalent formulation where the imaginary part of $\chi^o_{{\bf G}{\bf G}'}$ is calculated according to\cite{arguprb94,ar01,sinajpcm15}
\begin{eqnarray}
{\rm Im}[\chi^o_{{\bf G}{\bf G}'}({\bf q},\omega)] &=& \frac{2}{S}\sum^{\rm 2BZ}_{\bf k} \sum^{\rm occ}_{n} \sum^{\rm unocc}_{n'}
\delta(\varepsilon_{n{\bf k}}-\varepsilon_{n'{\bf k}+{\bf q}}+\omega) \nonumber \\
&\times&\langle{\psi_{n{\bf k}}} |e^{-{\rm i}({\bf q}+{\bf G}){\bf r}}|
\psi_{n'{\bf k}+{\bf q}}\rangle \nonumber \\
&\times&\langle{\psi_{n'{\bf k}+{\bf q}}} |e^{{\rm i}({\bf q}+{\bf G}'){\bf r}}|\psi_{n{\bf k}}\rangle.
\label{chi0_im}
\end{eqnarray}
In present calculations of Im$[\chi^o_{{\bf G}{\bf G}'}({\bf q},\omega)]$ we employ our own code.\cite{sichprb03} We replace the energy-conserving delta function in Eq. (\ref{chi0_im}) with a normalized Gaussian of width of 25 meV.
The real part of $\chi^o_{{\bf G}{\bf G}'}({\bf q},\omega)$ is obtained using the Hilbert transform over the energy range from 0 to 20 eV split into 10000 intervals.

Within time-dependent density-functional theory (TDDFT) formalism, the interacting density response function $\chi$ is obtained from the Dyson-like equation\cite{rugrprl84,pegoprl96}
\begin{widetext}
\begin{equation}
\chi_{{\bf G}{\bf G}'}({\bf q},\omega) = \chi^o_{{\bf G}{\bf G}'}({\bf q},\omega) + \sum_{{\bf G}_1,{\bf G}_2} \chi^o_{{\bf G}{\bf G}_1}({\bf q},\omega)[V_{{\bf G}_1}({\bf q})\delta_{{\bf G}_1{\bf G}_2}+K^{\rm xc}_{{\bf G}_1{\bf G}_2}({\bf q},\omega)] \chi_{{\bf G}_2{\bf G}'}({\bf q},\omega).
\label{Dyson}
\end{equation}
\end{widetext}

\begin{figure}[ht]%
\centering
\includegraphics[width=0.49\textwidth]{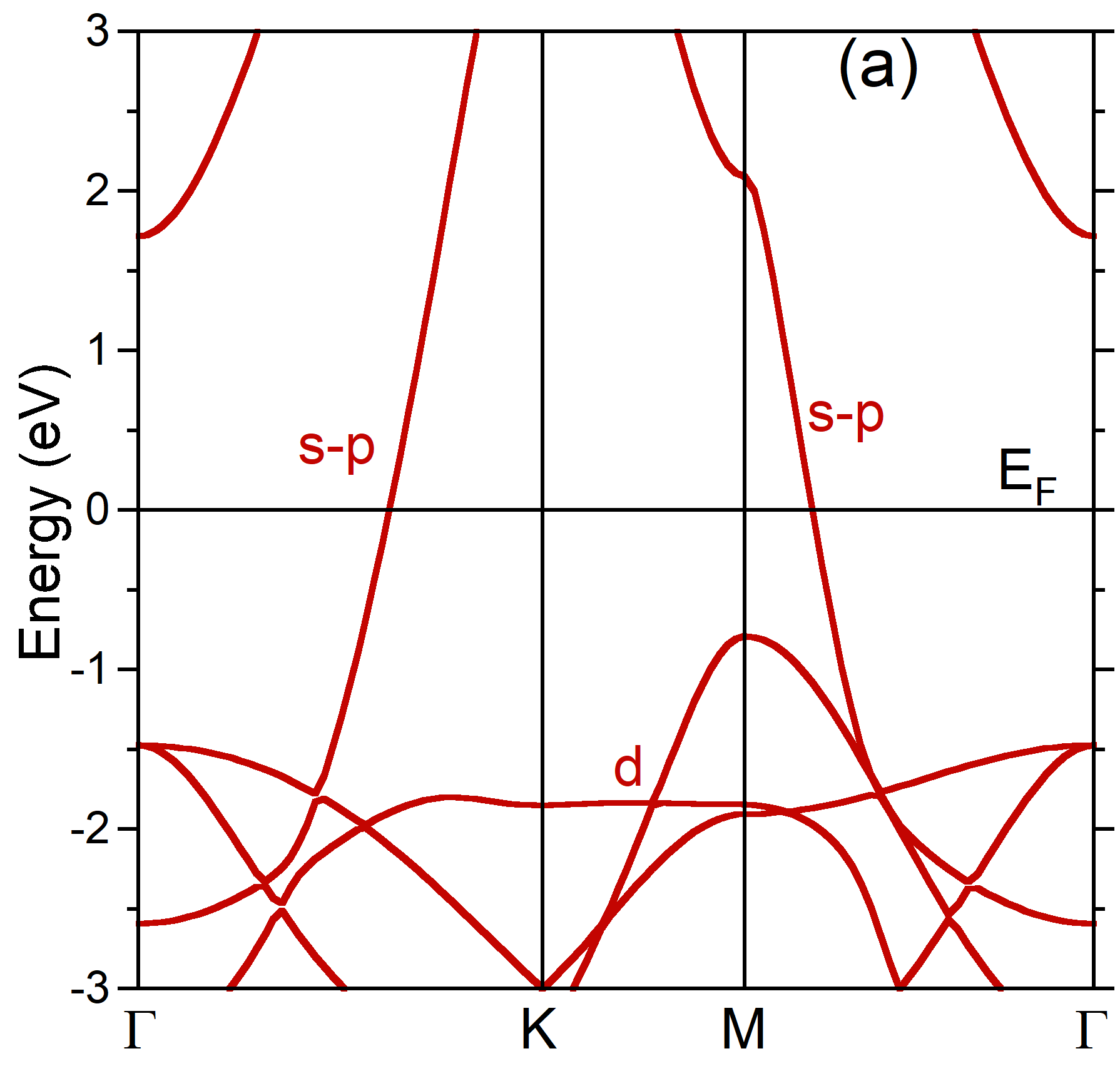}
\caption{Electronic band structure of a free-standing gold monolayer along the symmetry directions of the 2D Brillouin zone. The Fermi level, E$_F$, is set to zero.
}
\label{FIG-BS}
\end{figure}

For the interacting potential we employ a 2D Coulomb potential, $V_{\bf G}({\bf q})=2\pi/|{\bf q}+{\bf G}|$.
 The kernel $K^{\rm xc}$ accounts for the dynamical exchange-correlation effects. The present results are obtained within the random-phase approximation (RPA), when it is set to zero.

Here we obtain excitation spectra in the limiting case of a zero-thickness crystal.\cite{nanjp15}
The Fourier transform of the inverse dielectric matrix $\epsilon^{-1}_{{\bf G}{\bf G}'}({\bf q},\omega)$ is related to the interacting density-response function $\chi_{{\bf G}{\bf G}'}({\bf q},\omega)$ as
\begin{equation}
\epsilon^{-1}_{{\bf G}{\bf G}'}({\bf q},\omega) = \delta_{{\bf G}{\bf G}'} + \chi^o_{{\bf G}{\bf G}'}({\bf q},\omega) V_{{\bf G}'}({\bf q}).
\end{equation}
Inverting the $\epsilon^{-1}_{{\bf G}{\bf G}'}({\bf q},\omega)$ matrix, we obtain the macroscopic dielectric function $\epsilon^{M}({\bf q},\omega)\equiv \epsilon_1({\bf q},\omega) + {\rm i}\epsilon_2({\bf q},\omega)$, i.e. in the evaluation of the latter
the local-field effects\cite{adpr62,wipr63} (LFE) are included.

\section{Results and discussion}\label{sec2}

\subsection{Electronic structure}

\begin{figure}[h]%
\centering
\includegraphics[width=0.49\textwidth]{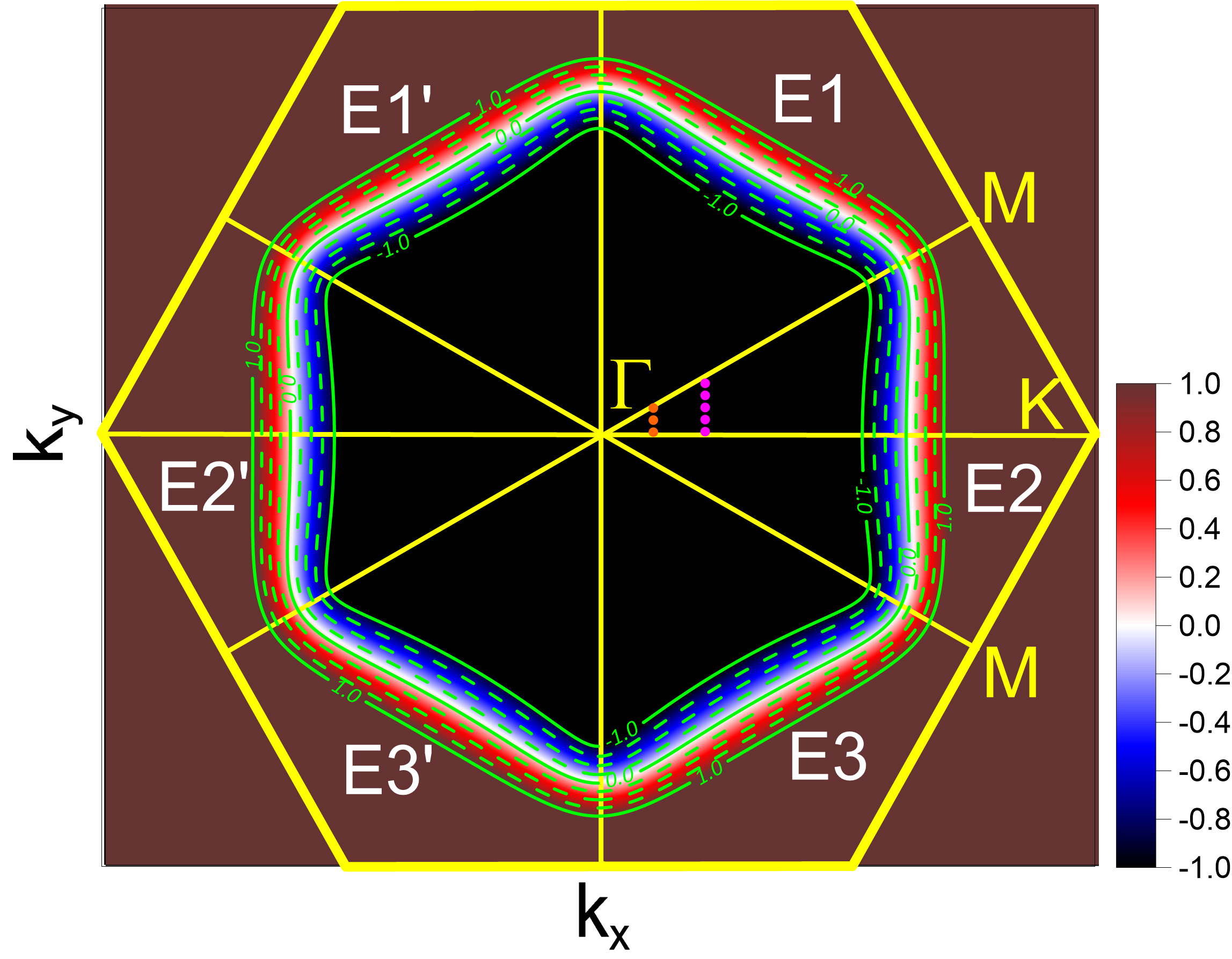}
\caption{
 Two-dimensional plot of the energy band dispersion over a whole 2D Brillouin zone in the -1 to 1 eV energy window at the Fermi level which set to zero.}
\label{FIG_FS}
\end{figure}

The calculated electronic structure of a free-standing Au(111) monolayer is presented in Fig. \ref{FIG-BS}. One can see that, like in bulk, a single band with a free-electron-like dispersion crosses the Fermi level. This band has mainly the $s-p$ character, although the proximity of the occupied 5$d$ valence states introduces some hybridization. The $d$-type  states reside at energies below $\sim$-0.8 eV. The calculated dispersion of the $s-p$ band along the $\Gamma K$ direction is very close to that of the Au energy band measured in a metallic state of a gold atomic layer intercalated between graphene and SiC substrate.\cite{folinc20}

Although the $s-p$ energy band presents a free-electron-like dispersion around the Fermi level, the interaction with the totally occupied valence $d$ states introduces some modifications in its dispersion. Indeed, it is rather far from to be isotropic, as one would expect in the free-electron-gas (FEG) model. To demonstrate this, in Fig. \ref{FIG_FS} we show the 2D map of its dispersion around the Fermi surface. One can see that the Fermi surface has strong hexagonal wrapping. As it will be shown in the following this fact has strong impact on the low-energy  excitation spectra of this system.

\subsection{Excitation spectra}

\begin{figure}
	\centering
	\includegraphics[width=0.99\linewidth]{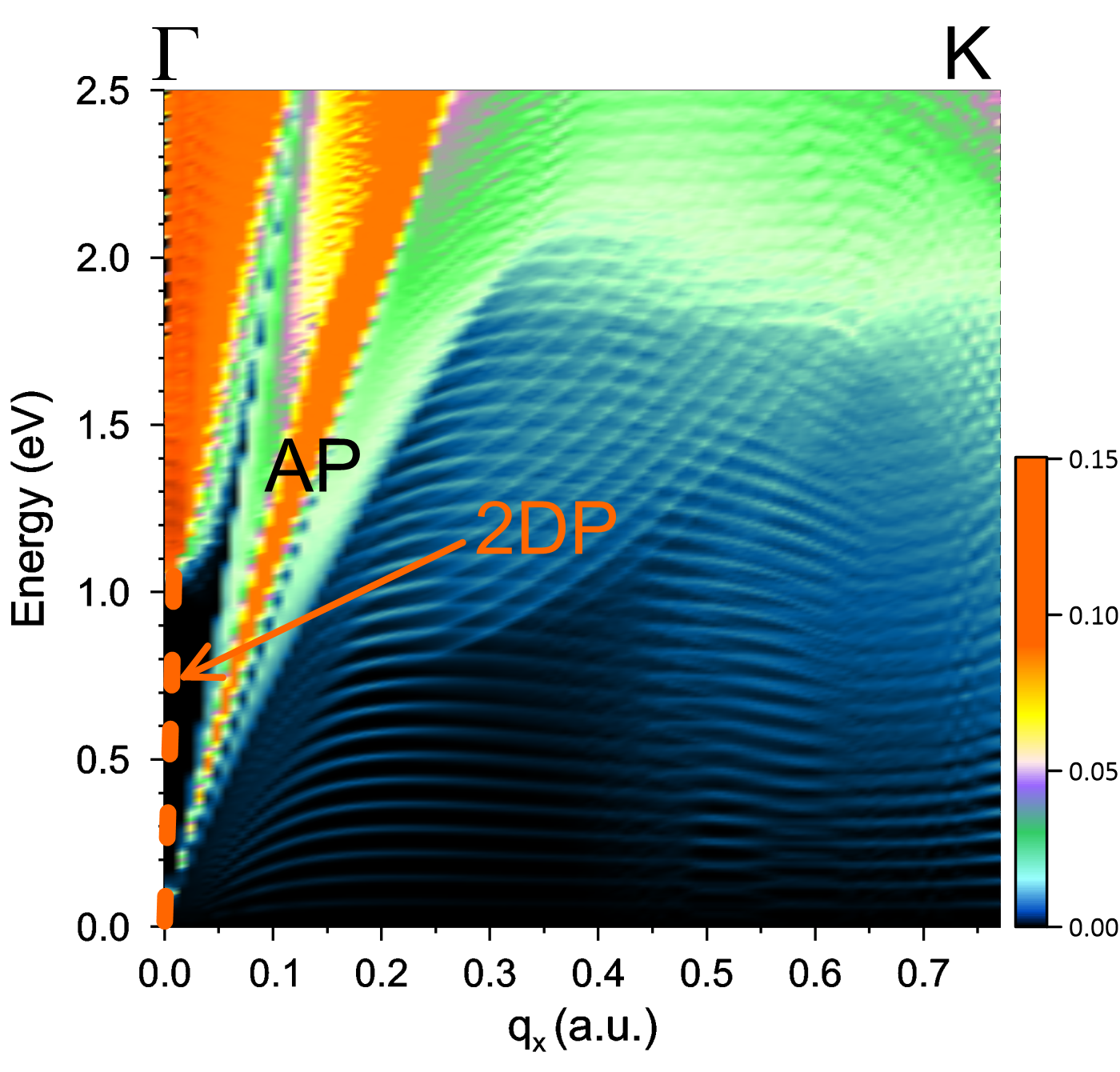}
	\caption{Loss function -Im$[\epsilon^{-1}({\bf q},\omega)]$ at momentum transfers ${\bf q}$ along the $[10]$ symmetry direction.
The dispersion of the conventional two-dimensional plasmon 2DP at small ${\bf q}$'s outside the e-h continuum is highlighted by dashed line. In this momentum region it has infinite lifetime.
The peak AP corresponds to the acoustic plasmon. The inter-band transitions involving the occupied $d$-type states are possible with energies higher than $\sim$1 eV when the 2DP peak acquires a finite width. Along this symmetry direction at finite momentum transfers these transitions become rather efficient at energies above $\sim$1.8 eV.
}
	\label{2D-LOSS_10}
\end{figure}

The  loss function of a gold ML evaluated at the momentum transfers ${\bf q}$ along the $[10]$ symmetry direction (coincides with the $\Gamma K$ direction of the 2BZ) is reported in Fig. \ref{2D-LOSS_10}.
The surprising observation is the presence of two peaks in the low-momentum region. Thus, in addition to the expected 2D plasmon branch marked as 2DP with a characteristic $\sim\sqrt{q}$ dispersion at vanishing $q$'s,\cite{stprl67} one observes at smaller energies a clear peak (AP) with a gap-less sound-like dispersion ($\omega_{AP}\sim q$).
The AP peak can be traced in an energy window from zero to energies as large as $\sim$2.5 eV. At larger energies this feature becomes rather broad due to efficient decay into the e-h pairs involving the occupied $d$-like electronic states. The presence of this peak in the excitation spectra contradicts our expectations, since  the system has only one energy band crossing the Fermi level.
In all previous studied cases, it was found that for existence of such an acoustic mode, the electronic system should have more than one group of carriers in {\it different} energy bands,\cite{picjp56} like subbands in the semiconductor heterostructures,\cite{picjp56,damaprb81,anformp82} surface and bulk states at metal\cite{sigaepl04,sipiprb05,dipon07,papaprl10,pomaprb11,jamuprb12,yajaprb12,piwejpcc14,vasmprl13,bebesr21} and nodal-line semimetal\cite{xuliprb26} surfaces, different energy bands in bulk metals,\cite{sibaprb09,sichprb09,cugaprb12,faarprb12,kascprb13,ecchprb19,cuwiprb21,gadinjp21} intercalated graphene,\cite{madeprb17,makaijms22}  silicene,\cite{gopiprb17,lishprb23}  germanene,\cite{shchsr17,pigosr22,lishprb23,liham25} borophene,\cite{tonoprb21} and two valley bands in graphene.\cite{denoprb13,pisinjp14,pisiprb16,gubaprb17,lishprb23}

\begin{figure}[ht]%
\centering
\includegraphics[width=0.49\textwidth]{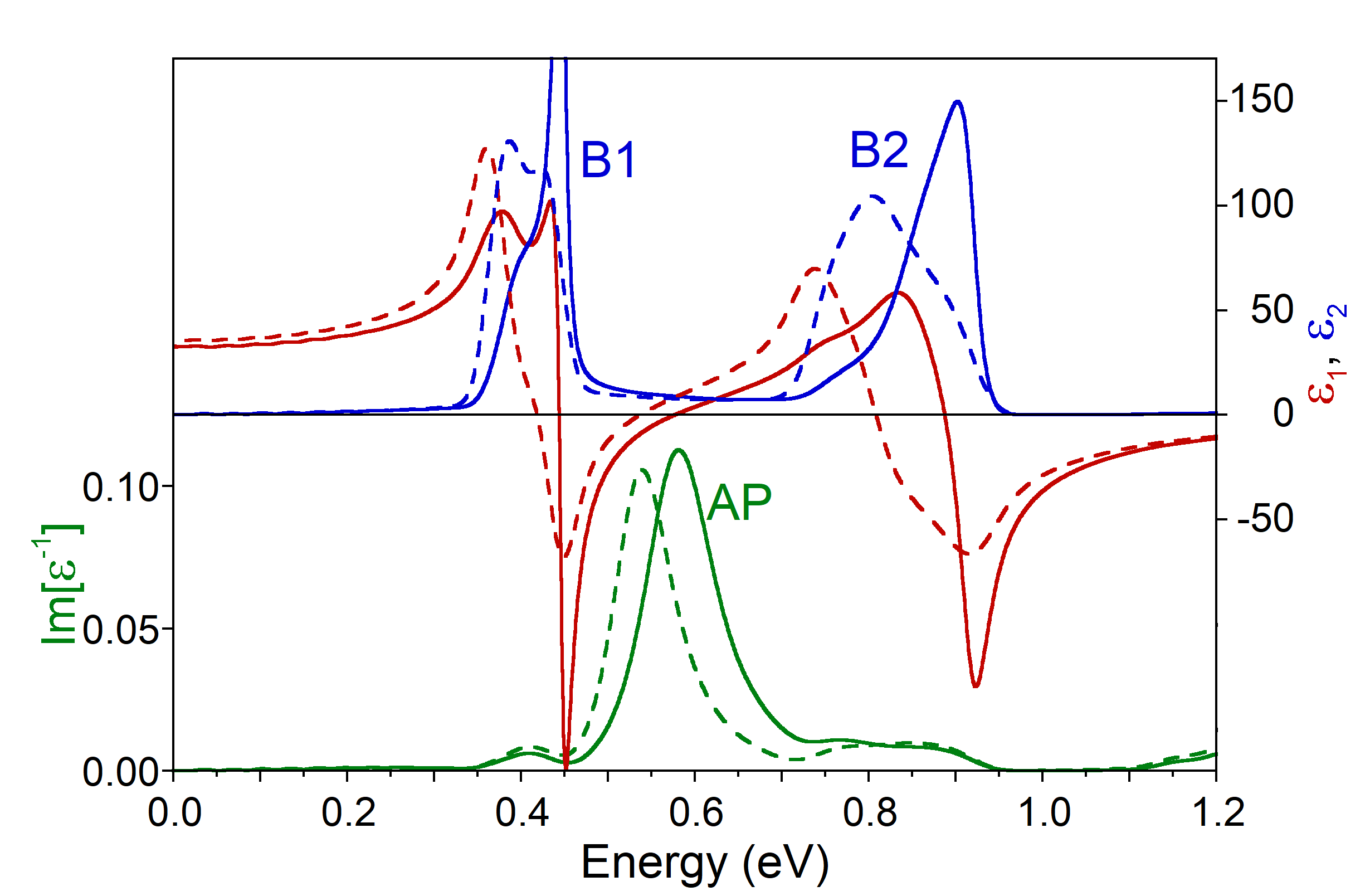}
\caption{Dielectric and loss functions evaluated at $q=0.048$ a.u. along the [10] symmetry direction.
The peaks B1 and B2 in the imaginary part of dielectric function $\epsilon_2$ are due to the intra-band transitions. The peak B1 is related to the transitions occurring at the sides E1 and E3 of the Fermi surface of Fig. \ref{FIG_FS} whereas the peak B2 is linked to the side E2. The dashed and solid lines stand for the results evaluated without and with inclusion of the LFE, respectively.
The peak in the loss function corresponding to the conventional 2D plasmon\cite{stprl67} is located at energies out of the scale of the figure.
}
\label{EPS_Q=012_000}
\end{figure}

In order to understand such an anomaly in the current case, in Fig. \ref{EPS_Q=012_000} we report the dielectric and loss functions evaluated at $q=0.048$ a.u. along the $[10]$ symmetry direction. The presented energy range corresponds to the intra-band excitations.
In the figure one can clearly see two dominating features B1 and B2 in $\epsilon_2$ signalling the presence of essentially two groups of carriers moving with different Fermi velocities. As a result, the microscopic (i.e. evaluated without inclusion of the LFE) $\epsilon_1$ crosses zero with a positive slope at the energy of $\sim$0.54 eV where $\epsilon_2$ presents a shallow minimum. Consequently, in the loss function obtained without LFE a well-defined peak (AP) appears at the energy of 0.54 eV signalling an existence of a well-defined collective charge excitation in the system. Inclusion of the LFE results in notable modifications in the dielectric function and the loss function. In particular, the peaks B1 and B2 in $\epsilon_2$  become more sharp and are blue-shifted. Consequently, the macroscopic $\epsilon_1$ crosses zero with positive slope at about 0.58 eV. Consequently, the energy position of the AP peak shifts upward by about 0.04 eV, as well.

As the energy dispersion of this mode is proportional to $q$ and gapless at $q \rightarrow 0$, we call it ''acoustic plasmon''.
One should distinguish this acoustic plasmon related to the Fermi surface geometry from a mode with a similar sound-like dispersion in the layered structures.\cite{grprb73,feap74,krmoprb88,bikrprb03,mahaprb08,beziprb24} The latter corresponds to the out-of-phase collective oscillation of the carriers in neighboring planes.
On the contrary, the nature of the mode found here is more close to a Pines' demon which corresponds to the out-of-phase collective oscillations of the carriers in {\it different} electronic energy bands.\cite{picjp56}

\begin{figure}[ht]%
\centering
\includegraphics[width=0.48\textwidth]{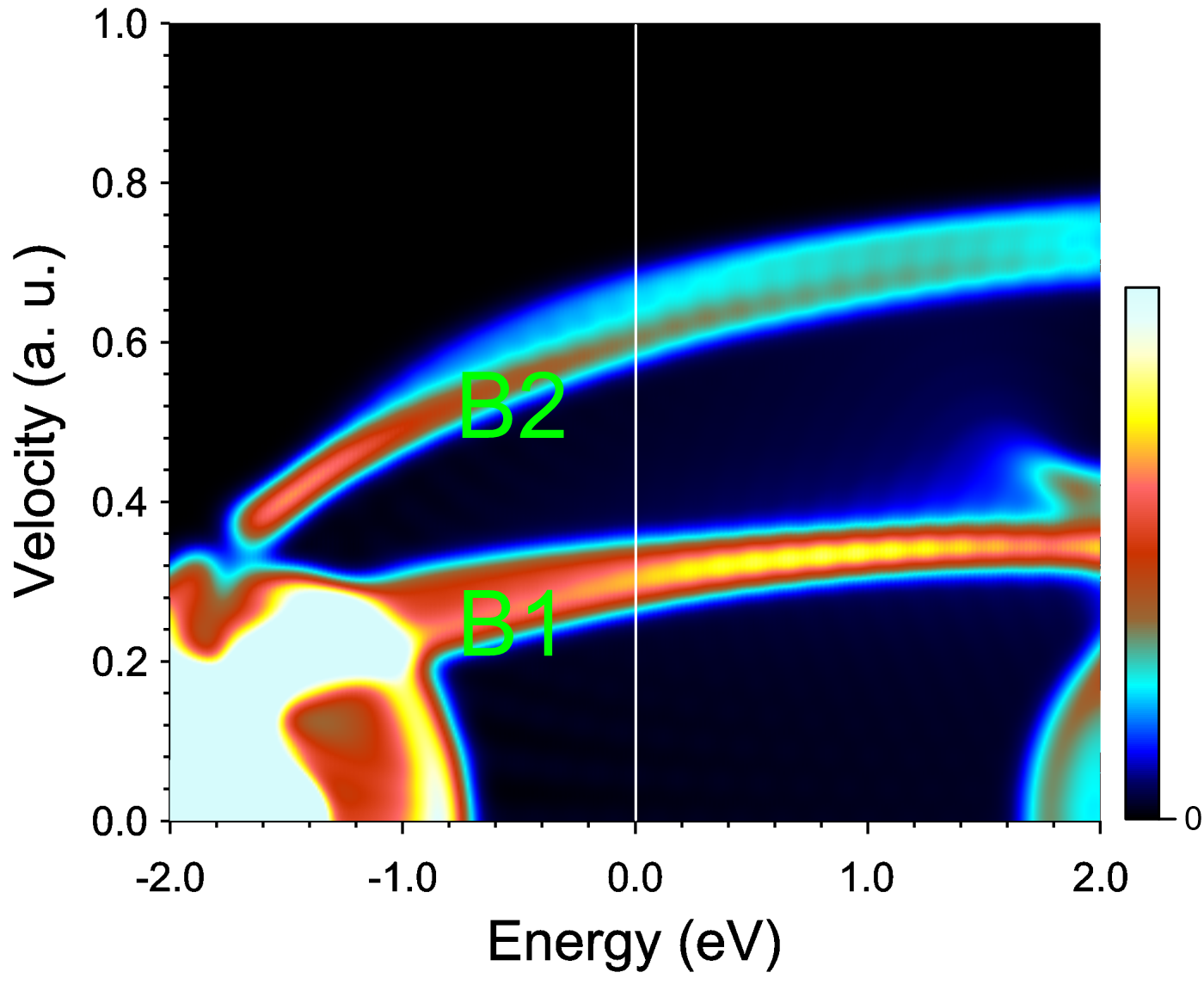}
\caption{Density of states versus energy and group velocity of the carriers moving in proximity to the Fermi surface in the [10] symmetry direction.
}
\label{DOS_10}
\end{figure}

To reveal why two groups of carriers exist at the Fermi surface in the Au(111) monolayer having only a {\it single} energy band crossing the Fermi level, in Fig. \ref{DOS_10} we show the distribution of the density of states (DOS) versus energy and the [10] component of the group velocity of the carriers around the Fermi surface. One can observe that in this DOS there are two dominating peaks B1 and B2 at the Fermi level. Such behavior is very different from the FEG case where the DOS presents a single peak at the velocity value close to the Fermi velocity.\cite{vasmprl13}

We rely a two-peak structure at $E_F$ in the DOS to the shape of the Fermi surface reported in Fig. \ref{FIG_FS}.
Remarkably, the shape of the Fermi surface of this band deviates strongly from a circular one which is a characteristic of FEG. Indeed, in the Au ML its shape is very close to the hexagon one.
Only small deviations from the almost straight lines can be detected at the centers of the sides. Moreover the vertices are slightly rounded. As a result of such a Fermi surface shape, the carriers at each side are moving with group velocities very close to a single Fermi velocity, ${\bf \upsilon_F}$, in the direction normal to this side. Hence, at a given in-plane momentum transfer, the carriers are split mainly in several groups with well defined group velocities. Notice, that indeed the slight deviation of the sides of the  Fermi surface from a straight line results in a slight splitting (hardly detectable in Fig. \ref{DOS_10}) of the B1 peak at the Fermi surface and its broadening. The splitting is seen more clearly in a two-peak structure of the B1 feature in the microscopic $\epsilon_2$ of Fig. \ref{EPS_Q=012_000}.

In the case of ${\bf q}$ pointing in the [10] direction, the peak B1 in $\epsilon_2$ of Fig. \ref{EPS_Q=012_000} is generated by the intra-band transitions at the sides E1 and E3. Its energy position in $\epsilon_2$ is proportional to the product of the Fermi velocity, ${\bf \upsilon_F}$, times $\cos(\pi/3)=1/2$. The peak B2 in $\epsilon_2$ is due to the transitions occurring in the vicinity of the side E2. Therefore, its energy position in $\epsilon_2$ is proportional to $\upsilon_F$, i.e., about 2 times larger that the energy of the B1 peak. As a consequence, the resulting AP excitation corresponds to the out-of-phase collective charge oscillations of the carriers at the side E2 forming one group and the sides E1 and E3 constituting the other one.
Hence, in the Au ML
 a two-component scenario, needed for existence of a second acoustic-like branch in the excitation spectra, is related to the hexagon shape of the energy band crossing the Fermi surface.
  
  Recently, such a scenario was predicted in the model calculations considering ideal geometries for the Fermi surface.\cite{musipccp22}
Also it was shown that similar modes can be realized in a normal state of high-T$_c$ cuprates\cite{sidrjpcl23,siefps25} caused by a specific shape of the Fermi surface.
Curiously, like in the cuprates case, almost all spectral weight in the loss function of Fig. \ref{EPS_Q=012_000} is concentrated in the AP peak. Only a small portion of the spectral weight corresponds to the e-h excitations. This is a remarkable example how dramatically can be reduced a number of the e-h excitations at the Fermi surface in a metallic system.

\begin{figure}
	\centering
	\includegraphics[width=0.99\linewidth]{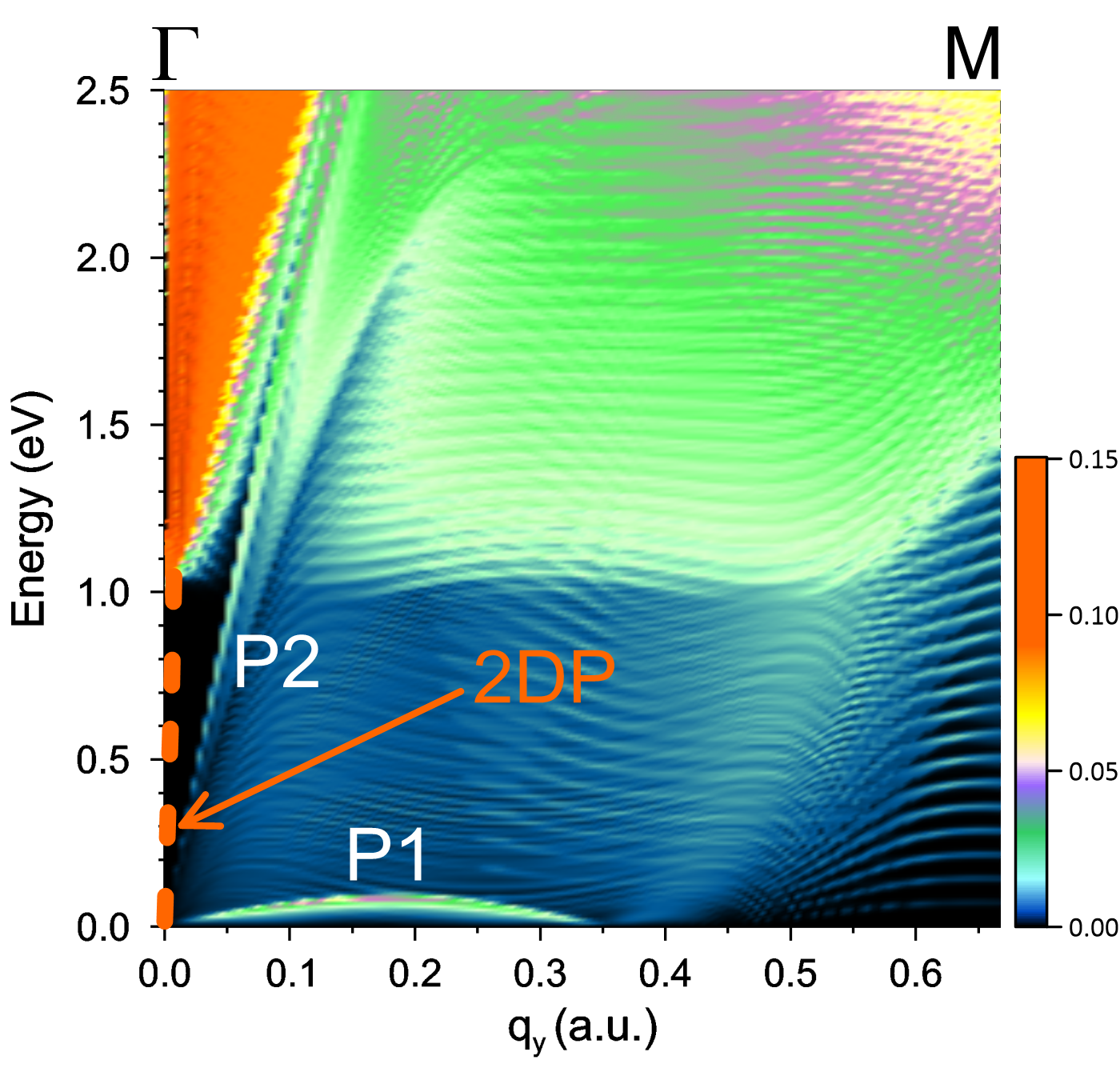}
	\caption{Loss function -Im$[\epsilon^{-1}({\bf q},\omega)]$ at momentum transfers ${\bf q}$ along the $[01]$ symmetry direction.
The dispersion of the conventional two-dimensional plasmon 2DP at small ${\bf q}$'s outside the e-h continuum is highlighted by dashed line. In this momentum region it has infinite lifetime.
The origin of the P1 and P2 peaks is explained in the main text. At energies above $\sim$1 eV the excitation spectrum is dominated by interband transitions involving the occupied $d$-type states and the 2DP peak acquires a finite width.
}
	\label{2D-LOSS_01}
\end{figure}

Figure \ref{2D-LOSS_01} exhibits the loss function versus energy and  momentum transfers ${\bf q}$ along the $[01]$ symmetry direction. Here, in addition to the 2DP plasmon peak, we observe two rather sharp peaks P1 and P2. Similar to the AP peak in Fig. \ref{2D-LOSS_10}, the P2 one in Fig. \ref{2D-LOSS_01} presents a sound-like dispersion. However, the intensity of the P2 peak is notably smaller.
Additionally, in Fig. \ref{2D-LOSS_01} a sharp  feature marked as P1 is presented at the lower energies. The P1 peak dispersion at small momentum transfers has a sound-like dispersion. It reaches a maximum energy of $\sim$0.1 eV at $q_y$=0.17 a.u. At larger $q_y$'s the dispersion turns to be negative and, after reaching a zero energy at $q_y$=0.34 a.u. this feature disappears. The latter momentum value  is very close to the edge size of the Fermi surface in Fig. \ref{FIG_FS}.

\begin{figure}[ht]%
\centering
\includegraphics[width=0.49\textwidth]{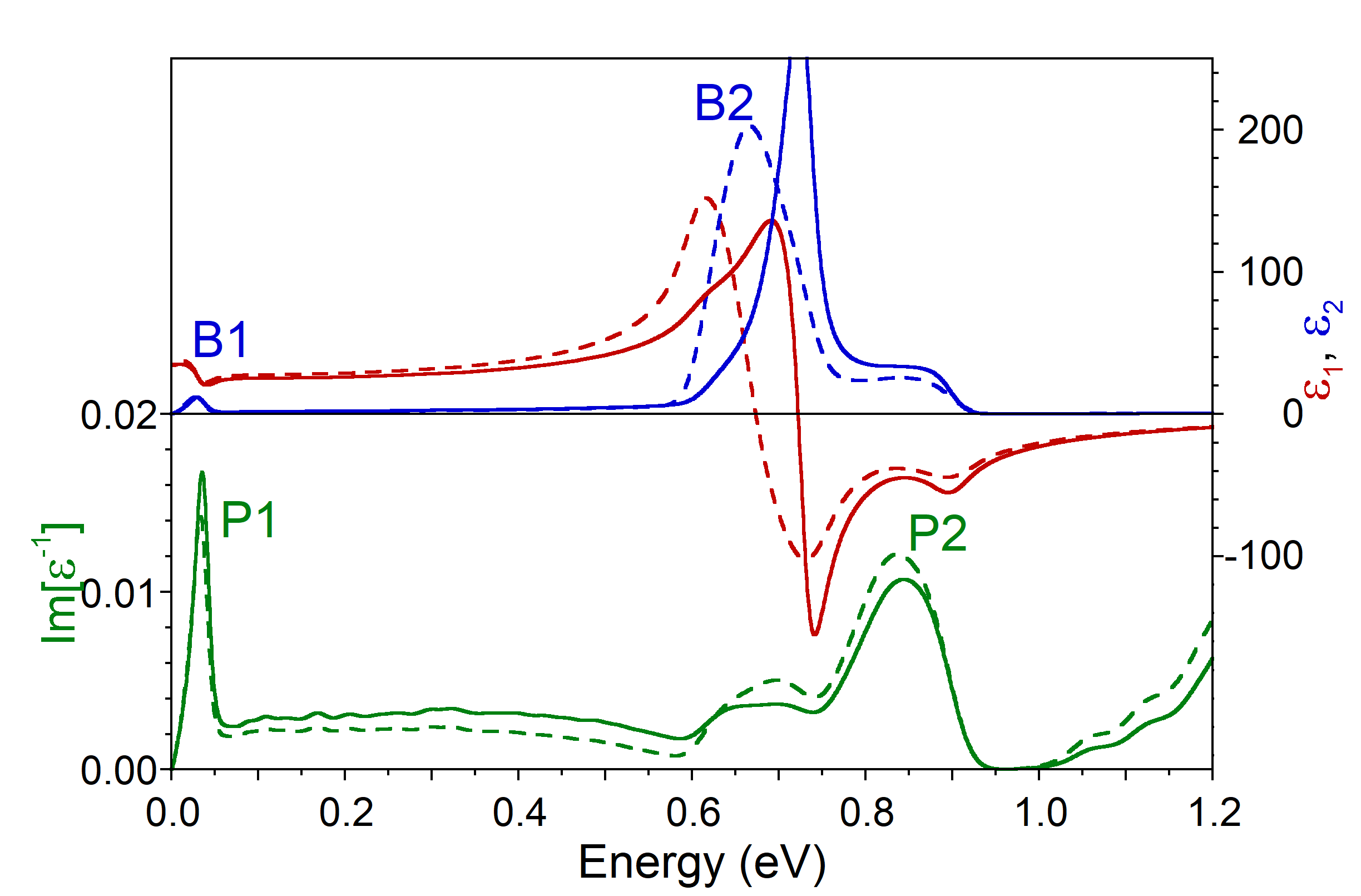}
\caption{Dielectric and loss functions evaluated at $q=0.046$ a.u. along the [01] symmetry direction.
The peaks B1 and B2 in the imaginary part of dielectric function $\epsilon_2$ are due to the intra-band transitions. The peak B1 is related to the transitions occurring at the sides E2 and E2' of the Fermi surface of Fig. \ref{FIG_FS} whereas the peak B2 is linked to the sides E1 and E1'. The dashed and solid lines stand for the results evaluated without and with inclusion of the LFE, respectively.
The peak in the loss function corresponding to the conventional 2D plasmon\cite{stprl67} is located at energies out of the scale of the figure.
}
\label{EPS_Q=000_020}
\end{figure}

In the loss function of Fig. \ref{2D-LOSS_01} the P2 peak has a conventional gap-less sound-like dispersion. This peak can be distinguished up to energies of about 2 eV. At larger energies it becomes rather weak and can be hardly detected due to a strong background produced by the inter-band single-particle excitations involving the occupied $d$ valence states.

To reveal the origin of the P1 and P2 features in the loss function of Fig. \ref{2D-LOSS_01}, we report  in Fig. \ref{EPS_Q=000_020} the dielectric and loss functions evaluated at ${\bf q}$=(0,0.046) a.u.
One can see that in this case, on the low-energy side, the peak B1 in the imaginary part of dielectric function $\epsilon_2$ stands at the energy of 0.029 eV. Consequently, the real part of dielectric function $\epsilon_1$ experiences strong modification in the nearby energy region. As a result of a such behavior of $\epsilon$, in the loss function a clear peak P1 emerges at the energy of 0.036 eV. Even though at this energy $\epsilon_1$ does not reach zero, we interpret the P1 peak as some kind of a collective electronic excitation, based on the fact that this peak in the loss function occurs at the energy where $\epsilon_2$ has a local minimum and $\epsilon_1$ maximally approaches zero. On the other hand, the linewidth of the peak P1 is of a similar magnitude as its energy, signalling the strong damping of this mode. Nevertheless, the presence of a clear peak P1 in the loss function may be interpreted as a signature of a second component in addition to the renormalized electronic system.\cite{dokirmp81} From Fig. \ref{EPS_Q=000_020} it is clear that the LFE have negligible influence on the properties of this mode.

\begin{figure}[ht]%
\centering
\includegraphics[width=0.48\textwidth]{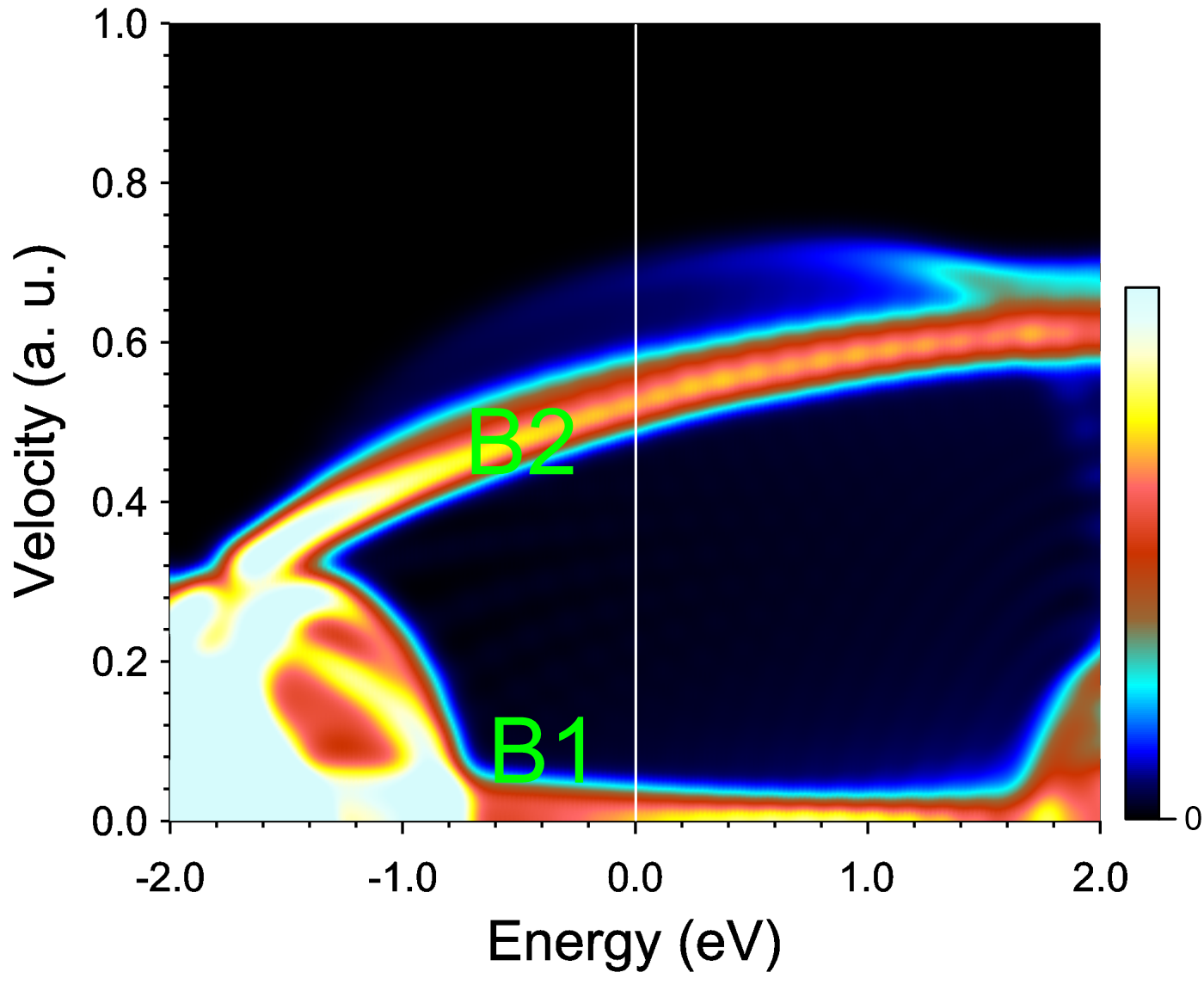}
\caption{Density of states versus energy and group velocity of the carriers moving in proximity to the Fermi level in the [01] symmetry direction.
}
\label{DOS_01}
\end{figure}

The origin of the B1 peak in $\epsilon_2$ can be traced to the peak B1 in the DOS reported in Fig. \ref{DOS_01}. This peak in the DOS is due to the electronic states residing at the edges E2 and E2' of the Fermi surface presented in Fig. \ref{FIG_FS}.
At the Fermi level this peak in the DOS has a maximum at $v_y\approx$0.02 a.u.

On the upper-energy side of Fig. \ref{EPS_Q=000_020}, the microscopic $\epsilon_2$ is dominated by a wide peak B2 centered at the energy of about 0.66 eV. On its right side a bump with much smaller intensity is centered at the energy of $\sim$0.85 eV.  This bump can be traced to the presence of a broad weak feature in the DOS of Fig. \ref{DOS_01} at the velocities above the B2 peak. Correspondingly, the departure of this feature from the peak B2 in the DOS is related to the deviation of the Fermi surface shape from the ideal hexagon. Despite of its smallness in comparison to the B2 peak, the presence of the bump in $\epsilon_2$ notably distorts $\epsilon_1$ at the neighboring energies. This  results in appearance  of a broad peak P2 in the loss function at 0.84 eV. Nevertheless, the peak P2 can be hardly interpreted as a well-defined collective charge excitation, since the amplitude of $\epsilon_1$ at this energy is even larger than in the P1 peak case. Moreover, $\epsilon_2$ does not present a remarkable minimum there. Nevertheless, following a general expression for a two-component plasma,\cite{dokirmp81} this peak can be classified as an overdamped mode.

In Fig. \ref{EPS_Q=000_020} one can observe that the LFE notably modify the dielectric function at energies in the vicinity of the B2 peak in $\epsilon_2$. It is seen that upon inclusion of LFE the B2 peak sharpens and its energy position is blue-shifted to $\approx$0.72 eV. As a result, the real part of dielectric function experiences strong modifications in this energy region as well. However, in spite of these facts, the loss function is hardly affected by LFE at these energies. Only small reduction in the intensity can be detected, which is compensated by small increase at energies below 0.6 eV.

\begin{figure}[ht]%
\centering
\includegraphics[width=0.49\textwidth]{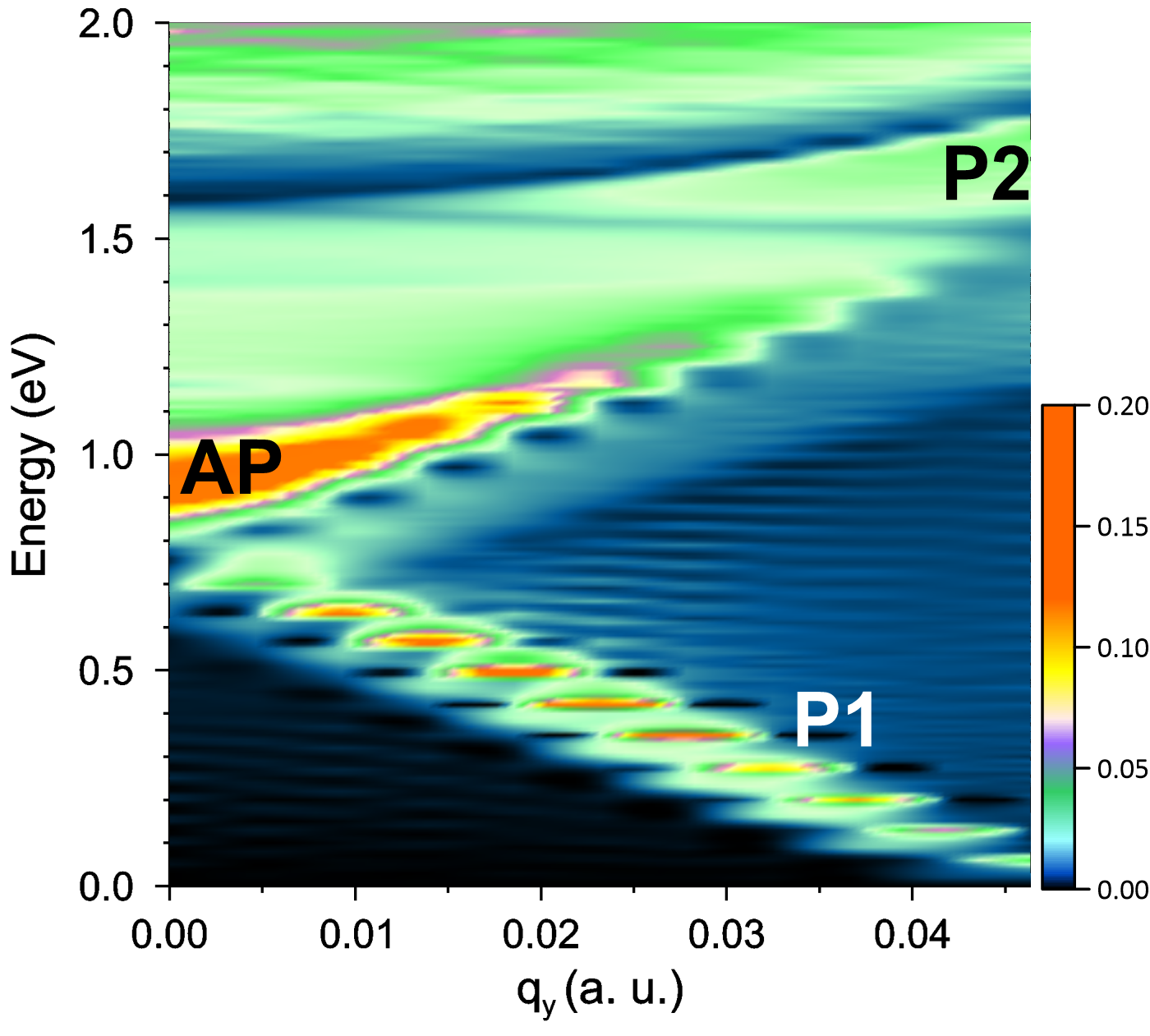}
\caption{Loss function -Im$[\epsilon^{-1}({\bf q},\omega)]$ at momentum transfers ${\bf q}$ along the $[10]$ direction at fixed $q_x=0.080$ a.u.
It is seen that at momentum transfers with finite $q_y$, additionally to a single AP peak existed along the $[10]$ symmetry direction (at $q_y=0$ in this figure), two other branches P1 and P2 exist. All these peaks are separated by regions with vanishing intensities.
 Discretization of the P1 branch and step-like structure in the case of the AP arise from the discrete 2BZ sampling. The horizontal width of the peaks in the P1 branch is determined by the distance between adjusting ${\bf q}$ points and does not reflect intrinsic momentum broadening.
}
\label{LOSS-2D_QX=020}
\end{figure}

Comparing  the loss functions reported in Figs. \ref{2D-LOSS_10} and \ref{2D-LOSS_01} for two symmetry directions of momentum transfer, one can observe that the 2DP plasmon has essentially isotropic dispersion. However the lower energy peaks differ in number.
In order to reveal whether these low-energy modes are related to each other, in Fig. \ref{LOSS-2D_QX=020} we show the loss function evaluated for momentum transfers along the line connecting the (0.080, 0) a.u. and (0.080,0.046) a.u. points as marked by orange dotted line in Fig. \ref{FIG_FS}. One can observe that the relation between all three peaks, AP, P1, and P2 is rather intricate. Apparently, they are connected at the ends of the momentum region.
In particular, upon departure from the $q_y=0$ point the AP peak disperses upward. Its intensity monotonically reduces. Eventually this peak reaches the $q_y$=0.046 a.u. point with a zero spectral weight. It looks like the AP peak and the upper-energy P2 peak are separated by an almost horizontal  valley located at energy of $\sim$1.55 eV. The P2 peak has similar upside dispersion with increasing $q_y$, although with much smaller slope. Also its intensity is significantly lower in comparison to the AP peak, as was pointed out previously, and gradually reduces upon departure from the $q_y$=0.046 a.u. point. The presence of the peak P2 can be detected at $q_y$ exceeding $\sim$0.01 a.u.

\begin{figure}[ht]%
\centering
\includegraphics[width=0.49\textwidth]{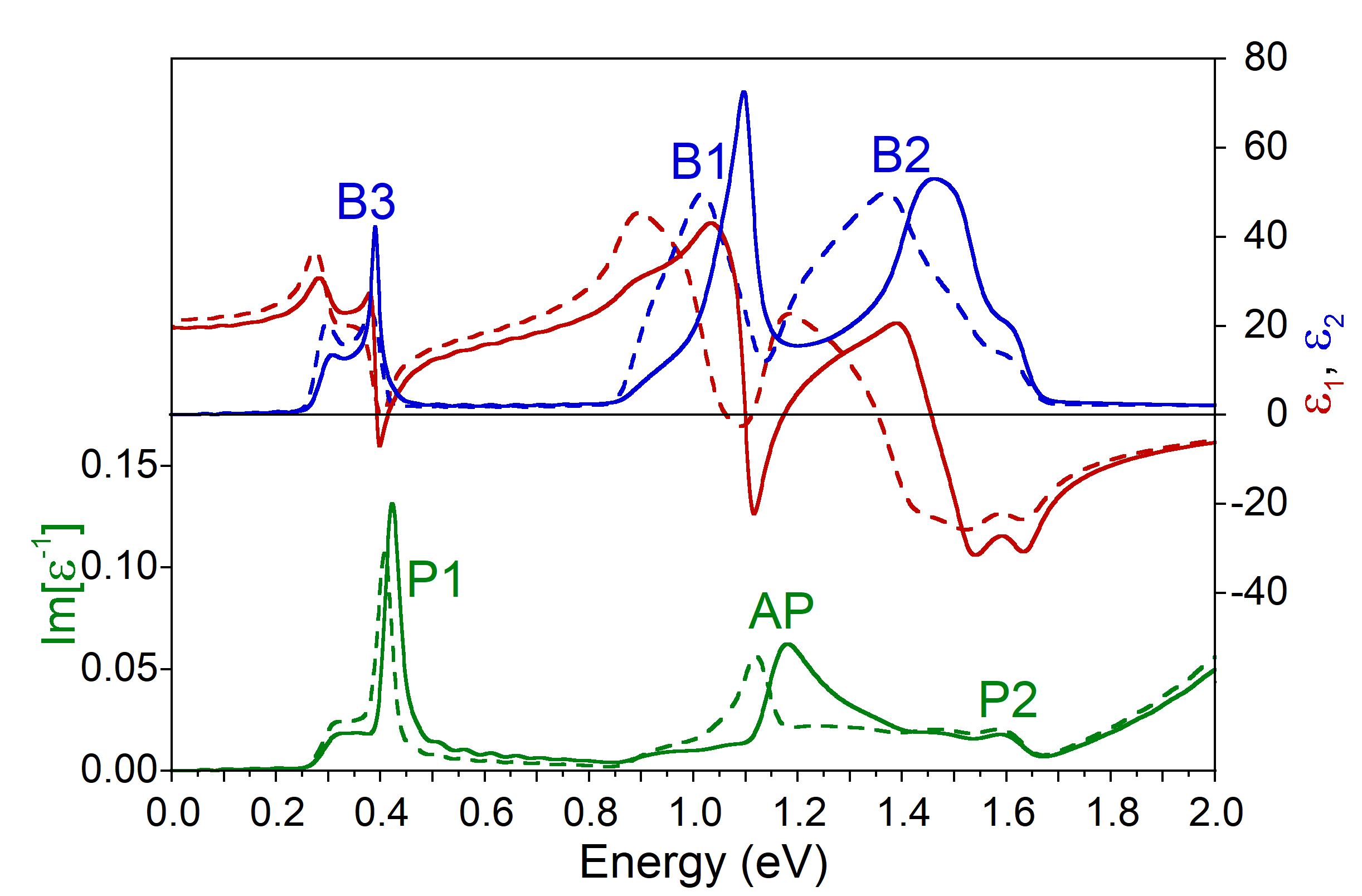}
\caption{Dielectric and loss functions evaluated at  ${\bf q}=(0.080, 0.023)$ a.u.
The peaks B1, B2, and B3 in the imaginary part of dielectric function $\epsilon_2$ are due to the intra-band transitions occurring at the sides E1, E2, and E3 of the Fermi surface of Fig. \ref{FIG_FS}, respectively.
The peak in the loss function corresponding to the conventional 2D plasmon\cite{stprl67} is located at energies out of the scale of the figure.
}
\label{EPS_Q=020_010}
\end{figure}

A more complex behaviour presents the lower-energy branch P1. At $q_y$=0.046 a.u. it has a small energy of about 0.06 eV. With reduction of the $q_y$ component of momentum the peak P1 disperses upward almost linearly. Initially the intensity of the P1 peak gradually increases. At momentum transfers of about 0.023 a.u. it reaches a maximum intensity. After that, at lower $q_y$'s its spectral weight gradually reduces and becomes zero approaching the $q_y=0$. Eventually, the P1 and AP peaks are separated by a valley with minimum intensity. Hence, in the case of the gold monolayer, the three clear peaks in the loss function present separate excitations, similar to the case of a normal state of high$-T_c$ cuprates.\cite{sidrjpcl23,siefps25} Nevertheless, apparently with variation of momentum transfer the intensities of these low-energy features are inter-related. The intensity of a given peak grows (decreases) at the expense (due to) of the nearby peak.

To explain the presence of three peaks in the loss function at ${\bf q}$ out of the symmetry lines of the 2BZ, in Fig. \ref{EPS_Q=020_010} we report the dielectric and loss
functions obtained at ${\bf q}$=(0.080,0.023) a.u. This momentum transfer corresponds to the center of the momentum region presented in Fig. \ref{LOSS-2D_QX=020}. One can detect immediately that at this ${\bf q}$
 the number of peaks in $\epsilon_2$ increases in comparison to the symmetric case presented in Fig. \ref{EPS_Q=012_000}. Similar to Fig. \ref{EPS_Q=012_000}, the peak B2 is generated by the transitions occurring at the side E2 of the Fermi surface. However due to reduced symmetry at this ${\bf q}$, the contributions to $\epsilon_2$ from  the sides E1 and E3 become different. In particular, the peak B1 in $\epsilon_2$ of Fig. \ref{EPS_Q=020_010} comes from the transitions at the side E1, whereas the two-peak feature B3 is produced by those occurring at the side E3. As a result, the real part of dielectric function $\epsilon_1$ experiences strong variations in this energy region with five zero crossings. In the energy regions where $\epsilon_1$ crosses zero with a positive slope, two clear peaks P1 and AP appear in the loss function signalling the presence of two charge collective excitations at energies of 0.42 eV and 1.17 eV, respectively. The lower-energy mode corresponds to the out-of-phase oscillations of the carriers at the sides E1 and E3. In the case of the AP plasmon the carriers at the sides E2 and E3 oscillate out-of-phase.

\begin{figure}[ht]%
\centering
\includegraphics[width=0.49\textwidth]{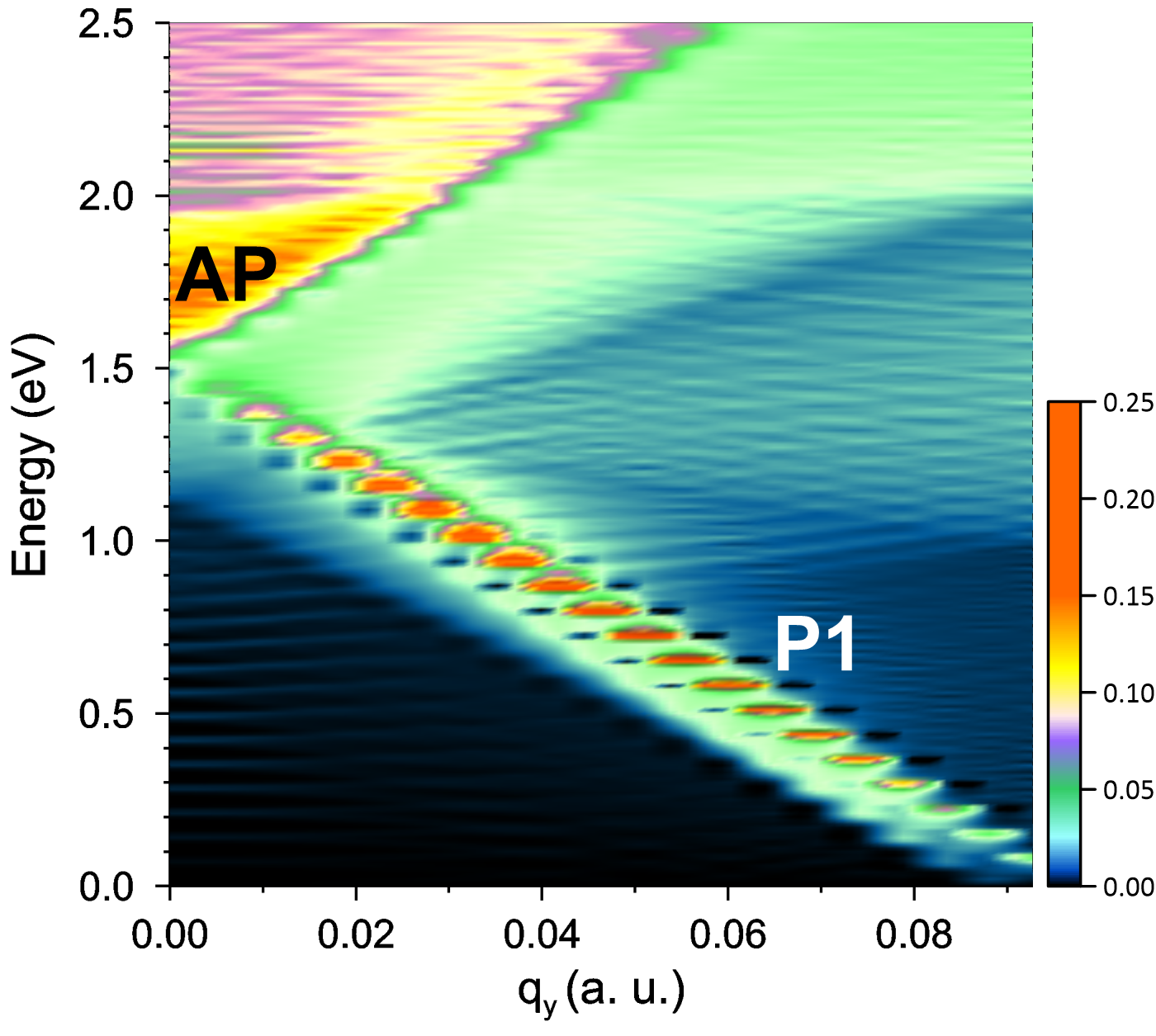}
\caption{Loss function -Im$[\epsilon^{-1}({\bf q},\omega)]$ at momentum transfers ${\bf q}$ along the $[10]$ direction at fixed $q_x=0.161$ a.u., as shown by orange dot line in Fig. \ref{FIG_FS}.
It is seen that the single AP peak existed along the $[10]$ symmetry direction (at $q_y=0$ in this figure) is separated from the branch P1 by a region with vanishing intensity.
 Discretization of the P1 branch and step-like structure in the case of the AP arise from the discrete 2BZ sampling. The horizontal width of the peaks in the P1 branch is determined by the distance between adjusting ${\bf q}$ points and does not reflect intrinsic momentum broadening.
}
\label{LOSS-2D_QX=040}
\end{figure}

Additionally, in Fig. \ref{EPS_Q=020_010} a weak peak P2 in the loss function can be detected at the energies around 1.6 eV. As it was explained previously, it has an origin in the bump
presented in $\epsilon_2$ in the same energy region accompanied by a distortion of $\epsilon_1$. It might be possible to interpret this peak as a strongly damped excitation where the carriers at the different places of the side E2 oscillate out-of-phase.

To confirm the tendency in the behaviour of the peaks in the loss function of Fig. \ref{LOSS-2D_QX=020}, in Fig. \ref{LOSS-2D_QX=040} we report the loss function evaluated at ${\bf q}$'s along the line connecting the momentum transfers
(0.161, 0) a.u. and (0.161,0.093) a.u. as marked by magenta dotted line in Fig. \ref{FIG_FS}. Here the P2 peak does not exist anymore due to efficient decay into the e-h pairs involving the occupied valence $d$ states. But, in the case of P1 and AP peaks we observe the same separation by the valley. The maximum intensity of the P1 peak is observed at $q_y\sim$0.046 a.u. The upper-energy AP peak gradually reduces its intensity upon departure from the $q_y$=0 point.

\begin{figure}
	\centering
	\includegraphics[width=0.99\linewidth]{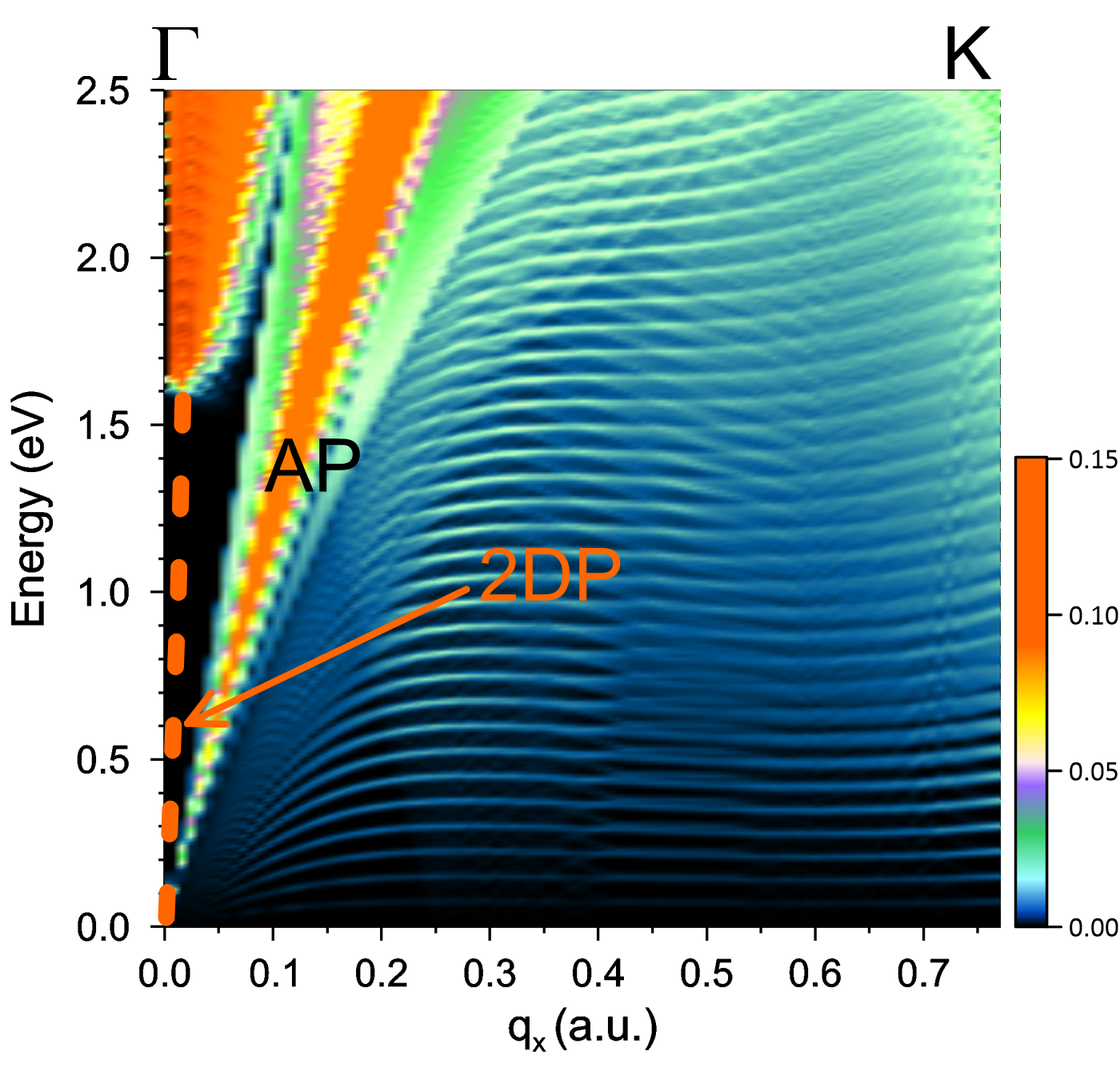}
	\caption{Loss function -Im$[\epsilon^{-1}({\bf q},\omega)]$ at momentum transfers ${\bf q}$ along the $[10]$ symmetry direction.
The dispersion of the conventional two-dimensional plasmon 2DP at small ${\bf q}$'s outside the e-h continuum is highlighted by dashed line. In this momentum region it has infinite lifetime.
The origin of the P1 and P2 peaks is explained in the main text. At energies above $\sim$1.6 eV the excitation spectrum is dominated by interband transitions involving the occupied $d$-type states and the 2DP peak acquires a finite width.
}
	\label{2D-DOP-LOSS_10}
\end{figure}

\begin{figure}
	\centering
	\includegraphics[width=0.99\linewidth]{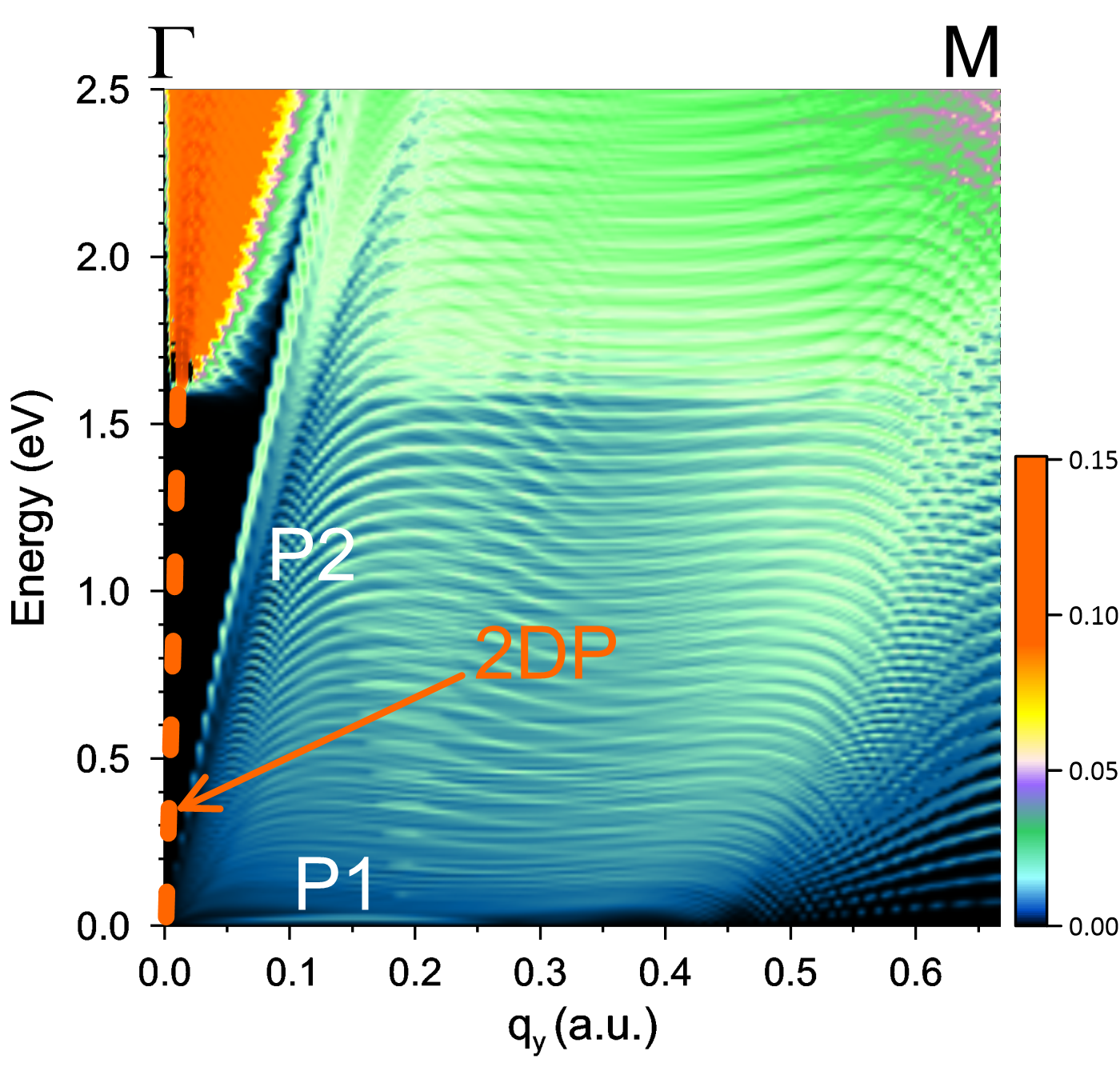}
	\caption{Loss function -Im$[\epsilon^{-1}({\bf q},\omega)]$ at momentum transfers ${\bf q}$ along the $[01]$ symmetry direction.
The dispersion of the conventional two-dimensional plasmon 2DP at small ${\bf q}$'s outside the e-h continuum is highlighted by dashed line. In this momentum region it has infinite lifetime.
The origin of the P1 and P2 peaks is explained in the main text. At energies above $\sim$1.6 eV the excitation spectrum is dominated by interband transitions involving the occupied $d$-type states and the 2DP peak acquires a finite width.
}
	\label{2D-DOP-LOSS_01}
\end{figure}

As it is stated above, we rely the remarkable dispersion of the P1 peak to the shape of the edges of the Fermi surface. Inspecting the Fermi surface reported in Fig. \ref{FIG_FS}, one can observe that it represents concave curvilinear hexagon. The isolines up to the energy of 0.7 eV have the same shape. At 0.7 eV  the isoline is almost flat along the sides. At higher energies  the shape of the constant-energy lines presents a convex curvilinear hexagon.
To demonstrate how the peaks in the loss function vary with doping in this energy range, we realized similar calculations of the dielectric and loss functions placing the Fermi level at the energy of 0.7 eV. This corresponds to the doping of the system with 0.17 electron.

Figure \ref{2D-DOP-LOSS_10} depicts a loss function calculated at this doping level at the momentum transfers aligned in the [10] symmetry direction. In general, the excitation spectrum is similar to the undoped case of Fig. \ref{2D-LOSS_10}. In particular, two plasmonic peaks are clearly seen on the left side of the figure. Nevertheless, some notable differences can be detected. Thus, in the doped case of Fig. \ref{2D-DOP-LOSS_10} the energy range where the 2DP mode is undamped is extended up to 1.6 eV. As for the AP peak, the slope of its dispersion is  affected very little by this doping. However, the broadening of this peak is lower in the case of Fig. \ref{2D-DOP-LOSS_10}. The other observation is the upward shift of the energy threshold for the e-h excitations involving the valence $d$ bands. In the doped case it is out of the energy scale of Fig. \ref{2D-DOP-LOSS_10}.

The loss function evaluated for the doped system at momentum transfers along the [01] axis is reported in Fig. \ref{2D-DOP-LOSS_01}. Three observations are similar to those in Fig. \ref{2D-DOP-LOSS_10}. Like there, the thresholds for the undamped 2DP peak and the interband e-h pairs excitations  involving the valence $d$ states are increased up to 1.6 eV. This is accompanied by narrowing of the P2 peak.  As for the lower-energy peak P1,  the situation there becomes rather intriguing. Its signature can be detected in the momentum range from zero to 0.28 a.u. But along this momentum interval its energy does not exceed a few meV, i.e. this mode becomes soft. The reduction of the maximum momentum where this mode is observed in comparison with the undoped case of Fig. \ref{2D-LOSS_01} is explained by visible increase of the radius of the rounded vertices of the isolines in Fig. \ref{FIG_FS} with the energy increase. This is accompanied by the reduction of the size of the flat regions of the sides.

Our finding are qualitatively different from the systems where strong deviation of the energy band dispersion around the Fermi level
from the FEG results in the deformation of the 2DP dispersion without appearance of additional lower-energy modes.\cite{haweprm21,nolyprb21,roiuprb25}

\section{Conclusions}\label{sec13}

In summary, our {\it ab initio} calculations show that in a gold monolayer the plasmonic structure in the low-energy domain governed by intra-band transitions is qualitatively different from the predictions of the free-electron-gas model. In particular, in addition to the conventional 2D plasmon with a characteristic $\sqrt{q}$ dispersion three other modes emerge at even lower energies. We trace the dispersion of all these additional modes having a sound-like dispersion in certain momentum directions and presenting strongly anisotropic in-plane behaviour. We rely the appearance of such anomalous modes to a nearly-hexagon-like shape of the Fermi surface of a single metallic energy band in the system. As a result, depending on the momentum direction, there exists a possibility to realize out-of-phase movement of the carriers at different sides of the Fermi surface with distinct well-defined Fermi velocities.

\section{Acknowledgments}\label{secA}

V.M.S. acknowledges financial support by Grant PID2022-139230NB-I00 funded by MCIN/AEI/10.13039/501100011033 and
The Department of Science, Universities and Innovation of the Basque Government, Grant No. IT1904-26.
The study of M.Yu.K.  was supported by the Program for basic research of the National Research University Higher School of Economics  and  the Ministry of Science and Higher
Education of the Russian Federation within the State assignment for the
Kapitza Institute for Physical Problems, Russian Academy of Sciences.


\end{document}